\documentclass[slac_one]{revtex4}
\usepackage{graphicx}
\usepackage{fancyhdr}
\def\met{\mbox{\ensuremath{\, \slash\kern-.6emE_{T}}}}
\def\pt{\ensuremath{p_T}}

\begin{document}

\title{Early measurements using W/Z in ATLAS} 

%

\author{D. Prieur}
\affiliation{STFC Rutherford Appleton Laboratory, Harwell Science \& Innovation Campus, Didcot, OX11 0QX, UK}

\author{On behalf of the ATLAS Collaboration}

\begin{abstract}
The LHC experiments are close to collecting collision data.
An overview of first physics measurements of the $W$ and $Z$ production cross-sections is presented. The electron and muon decay channels are considered. Emphasis will be given to data-driven approaches.
\end{abstract}

\maketitle

\thispagestyle{fancy}


\section{INTRODUCTION}

The study of the $W$ and $Z$ production is a fundamental area for the ATLAS \cite{AtlasPaper} early running. These processes are very well understood theoretically and with high expected counting rates they will provide stringent tests of QCD.
They will help to understand the detector performance and be used for: calibration of the electromagnetic calorimeter (EM), alignment of the muon spectrometer system, as well as for extracting the lepton identification efficiencies.
The first measurements will consist in determining the $W$/$Z$ and $W$/$Z$+jets cross-sections \cite{CSCnote}. Increased statistics will give access to fundamental electroweak parameters.
In this note, the emphasis is put on the early cross-section measurements assuming a LHC peak luminosity of ${\cal L}=10^{31}$~cm$^{-2}$s$^{-1}$.

\section{$W$/$Z$ CROSS-SECTION MEASUREMENT IN ELECTRON FINAL STATES}

\begin{figure}
\centering
\includegraphics[width=.61\textwidth]{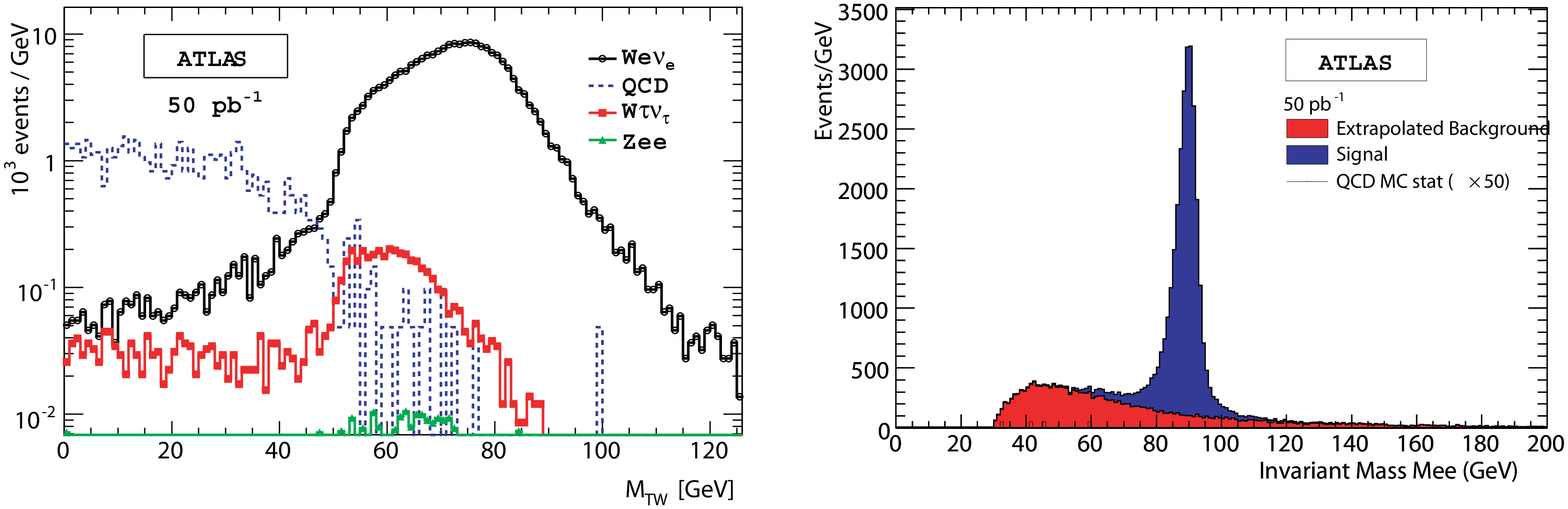}
\caption{\label{fig:mtelectrons} Left: Transverse mass distribution in the $W\to e\nu$ channel
  for signal and background, for $\int{\cal L}dt= 50$~pb$^{-1}$. Right: Di-electron invariant mass distribution in the $Z\to ee$ channel, for signal and background, for $\int{\cal L}dt= 50$~pb$^{-1}$.}
\end{figure}

The reconstruction of electrons is based on clusters in the EM calorimeter, with a matching track in the Inner Detector \cite{AtlasPaper}. The identification of isolated high-$p_T$ electrons is then based on the shapes of the EM showers, and on track reconstruction information. Three sets of identification criteria with different degrees of tightness (Loose, Medium, Tight) are used \cite{CSCnote}.
The selection of $W \rightarrow e\nu$ events proceeds as follows. The trigger selects events with at least one electron candidate with $E_{T}>20$~GeV. The analysis procedure selects events with exactly one electron candidate satisfying $E_{T}>25$ GeV, $\mid \eta \mid < 1.37$ or $1.52<\mid \eta \mid < 2.4$ and the Medium-type electron identification criteria. Then, the reconstructed missing transverse energy should satisfy $\met > 25$~GeV, and the transverse mass of the ($l,\nu$) system should satisfy $M_T>40$~GeV.
The resulting transverse mass distribution is shown in Fig.~\ref{fig:mtelectrons}.
Jet events constitute the largest background component. The jet production cross-section and fragmentation properties at the LHC are not well known and induce a significant uncertainty on the magnitude of this background.
A data-driven method to monitor the jet background is applied and is presented in section \ref{sec:datadriven}.
A MC study of this channel for $\int{\cal L}dt= 50$~pb$^{-1}$ gave a measured cross-section of $\sigma = 20530 \pm 40 \text{(stat)} \pm 1060 \text{(syst)} \pm 2050 \text{(lumi)}$~pb, to be compared with an input value  $\sigma = 20510$~pb \cite{PYTHIA, FEWZ}.

The analysis of $Z\rightarrow ee$ selects at least two electron candidates with $E_{T}>10$~GeV at the trigger level. The presence of two loosely identified isolated electrons with $E_T > 15$~GeV and $\mathrm |\eta|<2.4$ is then required. The resulting di-electron invariant mass distribution is shown in Fig.~\ref{fig:mtelectrons}.
As in the $W\to e\nu$ analysis, the jet background is estimated by a data-driven method.
The signal and background fractions are estimated simultaneously, via a fit to both
contributions.
The signal is described by the convolution of a Breit-Wigner and a Gaussian resolution function, and the background, completely dominated by jet events, by an exponential function.
A MC study of this channel for $\int{\cal L}dt= 50$~pb$^{-1}$ gave a measured cross-section of $\sigma = 2016 \pm 16 \text{(stat)} \pm72 \text{(syst)} \pm202 \text{(lumi)}$~pb, to be compared with an input value  $\sigma = 2015$~pb.

\section{$W$/$Z$ CROSS-SECTION MEASUREMENT IN MUON FINAL STATES}

\begin{figure}
\centering
\includegraphics[width=.61\textwidth]{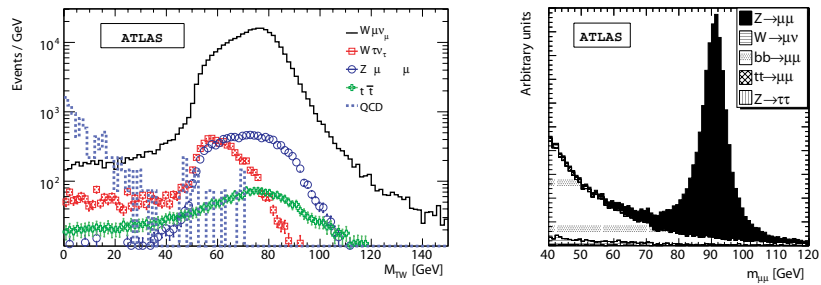}
\caption{\label{fig:mtmuons} Left: Transverse mass distribution in the $W\to\mu\nu$ channel,
  for signal and background, for $\int{\cal L}dt= 50$~pb$^{-1}$. Right: Di-muon invariant mass distribution in the $Z\to\mu\mu$ channel, for signal and background, for $\int{\cal L}dt= 50$~pb$^{-1}$.}
\end{figure}

The analysis of $W\to\mu\nu$ selects events with at least one muon candidate with $E_{T}>20$~GeV at the trigger level. The events are further selected requiring exactly one muon track candidate, identified in the muon and inner detector tracking system, satisfying $|\eta|<2.5$ and $p_T>25$~GeV. The energy deposited in the calorimeter, in a cone of radius $\Delta R = 0.4$ along the muon track, must not exceed $5$~GeV. In addition the event has to satisfy $\met >25$~GeV and $M_T>40$~GeV.
Figure \ref{fig:mtmuons} shows the corresponding $W$ transverse mass distribution before the $M_T$ cut.
In contrast to the electron channels, the jet background is less
important here.
The dominant backgrounds come from $W\to\tau\nu$ and $Z\to\mu\mu$
events. These processes are well understood theoretically
and can be safely estimated based on simulation.
A MC study of this channel for $\int{\cal L}dt= 50$~pb$^{-1}$ gave a measured cross-section of $\sigma = 20530 \pm 40 \text{(stat)} \pm630 \text{(syst)} \pm2050 \text{(lumi)}$~pb, to be compared with an input value  $\sigma = 20510$~pb.

The $Z\rightarrow \mu \mu$ analysis uses the $10$ GeV single muon
trigger. The data sample is further reduced by requiring two offline tracks with opposite charges, in the muon spectrometer only, with $|\eta| < 2.5$ and $\pt>20$~GeV. The invariant mass of the muon pair
is required to fulfil $|$91.2 GeV-$M_{\mu\mu}|<$20 GeV.
The corresponding invariant mass distribution before the mass cut is shown in
Fig.~\ref{fig:mtmuons}.
In this channel, the dominant background originates from $t\bar{t}$
events.
The jet background is
expected to be smaller, but is theoretically not well known.
Other backgrounds are smaller, theoretically well known, and contribute negligibly to the
overall background uncertainty.
A MC study of this channel for $\int{\cal L}dt= 50$~pb$^{-1}$ gave a measured cross-section of $\sigma = 2016 \pm 16 \text{(stat)} \pm64 \text{(syst)} \pm202 \text{(lumi)}$~pb, to be compared with an input value  $\sigma = 2015$~pb.

\section{DATA-DRIVEN BACKGROUND ESTIMATION FOR $W \to e\nu$ \label{sec:datadriven}}

\begin{figure}
\centering
\includegraphics[width=.33\textwidth]{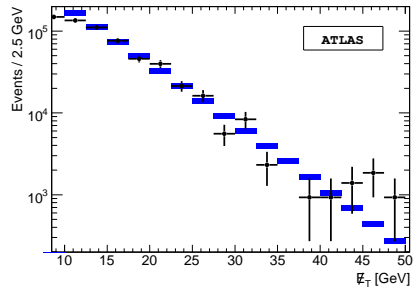}
\caption{\label{fig:datadriven} Comparison of the jet background (points with
error bars) and the fitted background (rectangles), for an integrated luminosity of   $\int{\cal L}dt= 50$~pb$^{-1}$.}
\end{figure}

The principle is to measure the normalisation and shape
of the jet background before the \met\ cut, in a
sufficiently pure jet sample. This sub-sample is then used to evaluate the rejection of the
\met\ cut, allowing a realistic estimation of the jet
background in the $W\to e\nu$ selection.
The jet background control sample is selected using a single
photon trigger with $E_T > 20$~GeV, and subsequent calorimeter only based electron
identification. Simulation studies show that these selections provide a
sample almost entirely composed of jet events, even at high values of
\met, and that the shape of the \met\ distribution is identical,
within the statistical precision, to that of the jet background in the
$W\to e\nu$ sample (see Fig.~\ref{fig:datadriven}). Above $\met>10$~GeV, the slope can be described
with the convolution of an exponential and a second degree polynomial
function.
After the subtraction of the estimated background from the signal
sample, the analysis then proceeds by applying the \met\ selection mentioned above.

\section{$W$/$Z$ + JETS}

The production of $W$/$Z$+jets events is an interesting measurement in itself.
In addition this process is
a background to many other Standard Model and beyond Standard Model physics channels. Furthermore these channels will test jet reconstruction techniques. Compared to the $W$/$Z$ inclusive production, more statistics is needed and the analysis done is based on an integrated luminosity of $1$~fb$^{-1}$. Selections are similar to the inclusive $W$/$Z$ production, except that one, two or three jets with $E_T>40$~GeV are required. At larger jet multiplicities, the dominant background arises from top quark events. The jet energy scale is the largest source of systematic error on the cross section. The initial uncertainty on the jet energy scale is expected to be 5-10\%. A jet energy scale with precision better than $10$~\% is required to distinguish between the LO/NLO predictions of the different Monte-Carlo generators.

\section{CONCLUSIONS}
The prospects for the measurement of the $W$ and $Z$ boson cross-sections in ATLAS have been presented. In the four channels ($W \rightarrow e\nu$, $Z\to ee$, $W\to\mu\nu$, $Z\to\mu\mu$) considered, high purity samples have been achieved after standard selections (high $p_T$ lepton identification, isolation and \met\ in
the $W$ final states) \cite{CSCnote}. The jet background has been controlled and subtracted using data driven methods.

\end{document}